\documentclass[12pt]{article}
\usepackage{scicite}
\usepackage{times}
\usepackage{amsmath,amssymb,graphicx,fullpage,color,mathtools,amsthm,xcolor}
\usepackage{scicite}
\usepackage{times}
 
\newenvironment{sciabstract}{
 \begin{quote} \bf}
 {\end{quote}}

\title{Active entanglement enables stochastic, topological grasping} 

\author
{Kaitlyn Becker,$^{1,2}$ Clark Teeple,$^{1}$ Nicholas Charles,$^{1}$ Yeonsu Jung,\\
$^{1}$ Daniel Baum,$^{3}$, James C. Weaver, $^{1,4}$ L. Mahadevan,$^{1,5\ast}$ Robert Wood$^{1\ast}$
\\
\footnotesize{$^{1}$School of Engineering and Applied Sciences, Harvard University, Cambridge, MA, USA}\\
\footnotesize{$^{2}$Department of Mechanical Engineering, Massachusetts Institute of Technology,Cambridge, MA, USA}\\
\footnotesize{$^{3}$Department of Visual and Data-Centric Computing, Zuse Institute Berlin, Berlin, DE}\\
\footnotesize{$^{4}$Wyss Institute for Biologically Inspired Engineering, Harvard University,Cambridge, MA, USA}\\
\footnotesize{$^{5}$Departments of Physics, and Organismic and Evolutionary Biology, Harvard University, Cambridge, MA, USA}\\
\\
\footnotesize{$^\ast$To whom correspondence should be addressed;
E-mail:  rjwood@g.harvard.edu,  lmahadev@g.harvard.edu}
}
 
\begin{document} 
\baselineskip20pt
\maketitle 

\begin{sciabstract}
Grasping, in both biological and engineered mechanisms, can be highly sensitive to the gripper and object morphology, as well as perception, and motion planning. Here we circumvent the need for  feedback or precise planning by using an array of fluidically-actuated slender hollow elastomeric filaments to actively entangle with objects that vary in geometric and topological complexity. The resulting stochastic interactions enable a unique soft and conformable grasping strategy across a range of target objects that vary in size, weight, and shape. We experimentally evaluate the grasping performance of our strategy, and use a computational framework for the collective mechanics of flexible filaments in contact with complex objects to explain our findings. Overall, our study highlights how active collective entanglement of a filament array via an uncontrolled, spatially distributed scheme provides new options for soft, adaptable grasping.
\end{sciabstract}

Securely grasping an object typically requires some knowledge of its size, shape, and mechanical properties. In the natural world, this is done, seemingly without effort, by elephants whose trunks can pick up a peanut or uproot a tree, orangutans whose combination of reaching and grasping allows them to brachiate rapidly in a complex arboreal environment, or a jellyfish whose tentacles collect stunned prey \cite{pfeifer2007,FLEAGLE1974,Madin1988}. In the engineered world of robotic grasping, much work has focused on understanding how to design the mechanics and dynamics of gripper hardware, as well as how to control such devices to interact with objects in the desired way. The form and stiffness of the grasper (relative to that of the target) determines the number (topology), shape (geometry), and magnitude (mechanics) of contacts and associated stresses on target objects \cite{mason2018toward, bicchi2000hands}. Inspired primarily by the remarkable dexterity of the human (and primate) hand, the majority of general-purpose graspers take the form of an articulated set of rigid links that are controlled locally, while an visual-motor feedback loop links perception, planning, sensing, and grasping actions \cite{cutkosky1986, bicchi1995closure}. Modern rigid grippers show great promise with many controllable degrees of freedom and embedded sensors \cite{jacobsen1986design, townsend2000barretthand, kochan2005shadow, bridgwater2012robonaut}, but can present challenges for grasp planning and control in the presence of uncertainty, or with complex target geometries \cite{mahler2017dexnet2, Morrison2020}. Some rigid grippers have incorporated sophisticated sensory and compliant contact surfaces to better perceive and adapt to target object with challenging geometries and mechanics\cite{She2021, kuppuswamy2020soft}. Alternatively, the introduction of strategic compliant elements into otherwise rigid fingers provides a form of mechanical intelligence that drastically reduces the planning and control requirements for successful grasping \cite{Dollar2010sdm, aukes2014design, friedl2018clash, catalano2014adaptive}.


An alternate strategy for robotic grasping is the use of soft actuators that avoid fine feedback control, and instead rely on mechanical deformation at multiple scales, both proximally and distally. Extending the concept of strategic compliance, soft grippers and hands devolve some of the mechanical complexity of a grasping task into morphology and passive mechanics of the gripper structure. This approach leads to conformable contact that, even in the absence of feedback control, is adaptable and robust to a range of variations in the target shape, size, and properties, and robust to damage in the hardware itself\cite{paoletti2017,shintake2018soft,rus2015design}. The multi-scale softness of all the gripper's surface can thus participate in a range of contact configurations with the target, even if the fingers are not independently controlled. Despite this added versatility that soft grippers offer, they are still typically constructed using a hand-centric design paradigm, where several digits are attached to a central hub \cite{deimel2016, zhou2018bcl, bhatt2021surprisingly}. In an alternative approach, high aspect ratio continuum actuators such as robotic tentacles \cite{mcmahan2006field, Martinez2013, Paek2015, Xie2020, Mazzolai2012, Giordano2021}, snakes\cite{Liao2020}, and tendrils \cite{Must2019} inspired by their biological counterparts\cite{gerbode2012}, leverage high aspect ratios and flexibility, to  adapt to a range of target objects. However, this still leaves open the question of how to grasp objects that are geometrically and topologically complex, and perhaps mechanically heterogeneous, such as a small potted plant, a fragile marine coral, or a branched artifact.


The distinction we make from previous works with soft continuum actuators and tentacles is the purposeful use of high aspect ratios soft actuators (filaments) arranged in arrays to achieve entanglement upon actuation of the filaments.  We introduce this qualitatively new topological grasping strategy inspired by the collective mechanics and dynamics of flexible filaments that can passively entangle on scales ranging from nanometric polymers to galactic magnetic flux lines. By engineering an array of pneumatically actuated filaments on the size scale of desired target objects, we realize a grasping strategy capable of adapting to the topological, geometric, and mechanical complexity of a range of target objects with minimal planning and no perception or feedback control. The basic building block of this strategy, shown in Figure~1A and 1B, is a slender elastomeric filament whose curvature can be modulated via pneumatic actuation. To become entangled and adapt to different grasping configurations, filaments must achieve sufficient curvature to bend and coil around each other and the target object, made possible by their slenderness. Coiling of our filaments is actuated via inflation of an off-center axial channel, which is sealed at one end. When an individual filament is actuated, it bends because of the eccentricity of the wall thickness, as shown in Figure~1B. This design enables the filament to deform into a highly curved state to form soft distributed contact zones either with a target object, itself, or other filaments as it reaches its operational pressure as shown in Figure~1C. These filaments are made following a dip-coating technique, where their characteristic curvature can be controlled by the position of the axial channel and their operating pressure can be tuned by the wall thickness~\cite{becker2020}. Furthermore, our dip molding methods (see supplementary text for details) allow for cheap, easy, and uniform construction of large arrays of actuators with a high aspect ratio (exceeding 200:1), giving filaments the compliance and sufficient length for entanglement. 

When combining filaments into arrays for grasping, a variety of design parameters can be chosen depending on the application requirements. For demonstration purposes, the configuration shown in Figure~1A and used in the experiments below uses twelve {300} {mm} long filaments connected to a single pressure source. Eight of the filaments are distributed evenly around a {50}{mm} diameter circle, and four of the filaments are evenly distributed around a {25}{mm} diameter concentric circle. (See supplementary text for further details). Not all of the twelve filaments will engage directly with an object for every grasp, as shown in Figure~1D, but the array can also be modified to incorporate more filaments (or a higher density of filaments) and increase the chance of entanglement. To demonstrate the topological complexity of structures for which we can achieve active collective entanglement using this approach, Figure~1E shows how, through filament self-entanglement, it is possible to cradle a spherical tennis ball, a bracket, and a branched tree (the latter of which are typically quite challenging for traditional grippers \cite{Morrison2020}). This soft adaptation, which occurs without perception, planning, or feedback, is due to the geometrically and topologically driven compliance of the actuated filament array as it interacts with complex objects.

 To quantify entanglement, we adapt ideas from knot theory and consider mesoscopic spherical volumes shown in Figure~2A that are larger than filament radii and smaller than the filament length and overall gripper dimensions. Within these volumes, we introduce the average crossing number - ACN (Fig.2B), defined as the average number of (unsigned) crossings of a filamentary structure over all possible viewing angles. The ACN provides an upper bound on the true (knot theoretic) topological complexity of filaments \cite{Buck2012}, so that we are guaranteed a conservative estimate of entanglement. Using the notation $e_{\alpha \beta}$ for the ACN between two open curves, $\alpha$ and $\beta$, we write:
\begin{equation}
    e_{\alpha \beta} = \frac{1}{2\pi} \int_\alpha \int_\beta {\frac{(\text{d}\mathbf{r}_i \times \text{d}\mathbf{r}_j) \cdot (\mathbf{r}_i - \mathbf{r}_j)}{|\mathbf{r}_i - \mathbf{r}_j|^3}}
    \label{eq:ACN}
\end{equation}
where $\mathbf{r}_i$ and $\mathbf{r}_j$ are the positions, and $\text{d}\mathbf{r}_i, \text{d}\mathbf{r}_j$ denote the infinitesimal vectors  along the centerline of $\alpha$ and $\beta$, respectively. For a pair of filaments,  the ACN will be smallest when the two filaments are parallel, and/or the filaments are far apart from each other, consistent with our intuition that such filament pairs are not entangled. By choosing a mesoscopic volume over which to evaluate the ACN that is larger than the filament radius but smaller than the system size, we can construct a scalar metric that provides a quantitative measure of entanglement.

We use the actuated filament-object configuration obtained with a CT scanner to deploy these metrics. In Figure~2C, we show the segmented and reconstructed scans of a set of structures (using the commercial software, Amira) in increasing order of relative entanglement corresponding to those shown in Figure~2A (and Fig.~1E), with colorized filaments for easy visualization. The filament center lines were used to compute the local spatial density of filaments, the local average crossing number, and the average crossing number of the entangled filament array within a ${20}~{mm}$ diameter spherical volume as described above. In Figure~2D,  we show the spatial density, and in Figure~2E, we show the local ACN (see supplementary text for further details). In Figure~2F, we show the contacts between the object and a filament colored in red, and contacts between filaments colored in blue (see supplementary text for magnified views).  As expected from the adaptive qualities of the filaments, the spatial organization of object contacts and inter-filament contacts changes with the topology of the target object. As compared to the simple sphere (tennis ball), the boundaries of the bar clamp and eight-branch tree grasps become more complex and represent a qualitative departure from traditional grasping \cite{mason2018toward}.


Collective entanglement of soft individual filaments eliminates the need for fine-grained planning and perception prior to grasping by distributing soft contacts across multiple filaments for greater cumulative engagement and entanglement with other filaments, the target object, or a combination of both. This strategy works particularly well in situations that are challenging for traditional soft and rigid grasping strategies, e.g., in grasping of topologically complex and delicate structures where traditional grasping notion of force closure used in deterministic grasping \cite{cutkosky1986, montana1992} is difficult to apply owing to the variability of contact number, size, shape and the corresponding contact forces.  

In order to compare entanglement-based grasping of target objects with more deterministic approaches, we use two experimental approaches, measuring the entanglement forces and grasp toughness, and measuring the grasp success when subject to positioning offsets (see supplementary text for more details of the methods). We define entanglement toughness as the energy required to pull an object out of the grasp, and measure it using our twelve-filament platform by attaching an object rigidly to the frame of an Instron universal testing machine and measuring the force-displacement curve till failure (see supplementary text for an image of the setup and details). For an operating pressure of ${172}~{kPa}$ (${25}~{psi}$), we find that the maximum grasping forces achieved over the various objects was ${27.6}~{N}$, which is comparable to many robotic hands with soft, pneumatic fingers operating at similar pressures \cite{shintake2018soft}. Toughness values for the entangling 12-filament gripper range from ${10}~{mJ}$ for a ${10}~{cm}$ sphere, to ${380}~{mJ}$ for a simple branched structure, and ${770}~{mJ}$ for a vertical ${51}~{mm}$ cylinder, consistent with our intuition that increasing object complexity (for the branched structure) and contact area (for the cylinder) increases the entanglement toughness. For comparison, values for the grasp toughness of recently developed soft grippers holding on to cylinders with diameters of ${51-76}~{mm}$ are ${200-700}~{mJ}$ (see supplementary text for details).

To evaluate the efficacy of entanglement in successfully grasping, lifting, and moving an object from its initial to its final position, we used sequences like those shown in Figure~3A with a sphere, a hollow cylinder, as well as four objects identified as part of an adversarial object set for robotic grasping \cite{mahler2017dexnet2} (See supplementary text for details on test objects). The approach trajectory in all cases consisted of draping the filament array over the target object from above (``top-drape'') using a robot arm (UR5e, Universal Robots), with 20 grasp trials per object, the results of which are shown in Figure~4B. We evaluated the entanglement gripper's sensitivity to positioning errors using grasps with controlled centering offsets in increments of ${10}~{mm}$, for 20 trials at each location. (See supplementary text for further testing information). The results of these experiments are shown in Figure~4C as a function of the offset between center axis of the gripper and the center axis of the target object (normalized to the object radius). Overall, we found that the entanglement gripper is tolerant to large centering errors (grasp success diminishes by less than 10\% for centering errors of $0.2 \times$ the object's radius or less), and particularly large errors for the eight-branch tree (less than 10\% reduction in grasp success for centering errors of $1.5 \times$ the object's radius or less). Overall, we find that our stochastic entanglement strategy works well for grasping topologically and geometrically complex objects but is less successful with simpler objects like spheres, cylinders, and cubes, which can be easily grasped using traditional rigid grippers \cite{mahler2017dexnet2, Morrison2020}.

The design space of active entanglement can be understood in terms of dimensional analysis. For a single filament of radius $r$, length $l$, internal channel radius r, channel eccentricity $\epsilon r, \epsilon \in [0,1]$, elastic modulus $E$, in an array with a characteristic spacing $d$, actuated by a pressure $p$, the design space of the gripper  is spanned by the following dimensionless parameters: gripper filament areal density $\phi_G = r^2/d^2 \ll 1$, a scaled pressure $p/E$, and finally the geometric arrangement of the filaments denoted by a scalar $S$. Additionally, if we also vary the filament length, internal channel radius, and eccentricity, we can control $l/r, \delta, \text{and}\ \epsilon$. Finally, moving from terrestrial to aquatic environments provides an additional parameter, $l/l_g$, where $l_g = (Er^2/\Delta \rho g)^{1/3}$ is a gravitational length, with $\Delta \rho$ being the difference in the density between the filament material and the ambient medium. Here, we will focus primarily on varying the gripper areal density of the filaments, $\phi_G$, for simplicity, recognizing that there is a vast range of possibilities for further exploration. An object to be grasped, on the other hand, can be characterized by its size, $R_T$, the topological complexity of its branching structure, which we capture in a simplified form using its effective volumetric density, $\phi_T$, within a convex hull around the object, and finally its mass density, $\rho_T$, that determines the object weight $\rho_T R_T^3 g$. The efficacy of the gripper is a function of its topological and geometrical complexity, as well as that of the target, and is a function of these dimensionless parameters. 

Upon activation, the characteristic curvature, $\kappa$, of a filament  subject to pressure $p$ scales as $\kappa \sim p(1-\delta)^2 \epsilon/rE$ and follows from a simple torque balance (see supplementary text for details). For grasping when gravitational effects can be neglected (e.g., in an aquatic environment, or when $l/l_g \ll 1$), the radius of curvature of a filament $R \sim \kappa^{-1}$ must be smaller than the overall size of the target $R_T$, and furthermore, the length of the filament $l$ must satisfy $l \ge R_T$  to enable distributed contact. This is a conservative estimate, since in an array of filaments of areal density $\phi_G$, it may be possible to collectively entangle with the target since the effective curvature of a tangle will scale as $\kappa f(\phi_G)$, where $f(\phi_G) \ge 1$ is a function that depends on the details of the filament array geometry. Therefore, a simple scaling relation for entanglement grasping via an array of long actuated filaments is given by $pR_T (1-\delta)^2 \epsilon f(\phi_G) /rE \ge 1$. In terrestrial environments, an additional condition is that the weight of the target must be supported by the entanglement, so that $\rho_T R_T^4 g  \le Er^3p(1-\delta)^2 \epsilon f(\phi_G)$, a scaling result that follows from the balance between elastic and gravitational torques. These two scaling estimates characterize the geometric and mechanical requirements for grasping. To explore these ideas, we tested filament spatial density by varying the filament spacing (Fig. 3D), object spatial density by varying the branch length of the eight-branch tree (Fig. 3E), and relative density of the object to the filaments by varying test tree mass (Fig. 3F). As expected, we see a drop in performance when the spatial density of the filaments or target branch length decreases, reducing the probability of entanglement. The largest tree branches also showed a decrease in robustness to normalized centering error, likely due to a combination of increasing mass, a greater offset between contact points and the center of mass, and effectively few accessible branches as the circumferential distance between branches. 

To go beyond the scaling ideas above, we use numerical simulations of a director-based Cosserat continuum framework for slender filamentous objects \cite{Cosserat1909,Gazzola2018}  to explore the mechanics of rods capable of bend, twist, stretch, and shear deformation modes, all necessary to follow the geometrically nonlinear deformations of our elastomeric filamentous actuators, including inter-filament contact, friction from sliding contact, gravity, and internal viscous dissipation \cite{Gazzola2018}.  The actuation of the filaments is accomplished by introducing an intrinsic curvature along the length of the filaments at the instant of actuation and assuming that the actuated shape equilibrates rapidly relative to the dynamics of entanglement or contact creation with the target. In a gravitational field,  the gripper filaments curl into helical structures and make contact with other filaments and the target, leading to a soft entangled grasp.  Although our simulation framework does not account for the effects of static friction or electrostatic forces due to charge build-up in sliding filaments, it is still capable of capturing the qualitative aspects of entanglement-mediated grasping, replicating our experimental observations (For details, see supplementary text). In Figures~3B-C, we show the ability of our simulation framework to tangle with and lift a branched structure, (the eight branch tree), remaining successful until the scaled positioning offset is as large as $30\%$ of the target size, a conservative estimate given that we have not accounted for frictional effects in the simulations. A side-by-side comparison of the experimental and simulated grasps are shown in Figure~1D (see also SI video 2).

Our simulations also allow us to explore the phase space spanned by the ratio of the target object spatial density $\phi_T$, the filament spatial density $\phi_G$, and a ratio of the density of the target object to that of the gripper, shown in Figure~3G, with each point being the result of seven simulations, shown along with the results of physical testing by varying filament spacing, branch length, and eight-branch tree mass corresponding to Figure~3D-F. The contour plots show the success rate (defined by the ability of the gripper to maintain a grasp for one minute). The simulated phase space shows a slight underestimate of the performance achieved in physical testing, likely because the effects of static friction were not accounted for.

Secure grasping of an object in both animate (human) and inanimate (robotic) settings requires a characterization of the size, shape, mass distribution, and stiffness of the target, and suggests crucial roles for perception, planning, and action with feedback. Here we demonstrate an embodied solution to this problem, relying on the active entanglement of an array of slender, pneumatically actuated filaments for adaptable grasping without perception, planning, or feedback. Our gripper can entangle, wrap, or cradle target objects via distributed soft contacts, and pick up targets with a range of sizes, topological complexities, geometric shapes. A scaling and computational framework for entangling thin elastic filaments corroborates our experimental observations. All together, our approach to the problem of robotic grasping complements traditional solutions by replacing grippers with few degrees of freedom but complex feedback control strategies with infinite-dimensional compliant filaments that are morphologically complex, but operate without feedback. This ability to use complex morphology (geometry and topology) and dynamics (physics) with simple control will expand the range of objects conducive to robotic grasping.

\bibliography{nembib}
\bibliographystyle{Science}

\section*{Acknowledgments}
This work was supported by a grant from the Office of Naval Research (Award \#N00014-17-1- 2063), grants from the National Science Foundation (Awards \#EFRI-1830901, \#DMR-1922321, \#DMR-2011754, \#DBI-1556164, and \#EFMA-1830901), the National Science Foundation Graduate Research Fellowship (under Grants \#DGE1144152 and \#DGE1745303,), the Wyss Institute for Biologically Inspired Engineering, the Simons Foundation and the Henri Seydoux Fund.


\clearpage

\begin{figure}[t!]
\centering
\includegraphics[width=\textwidth]{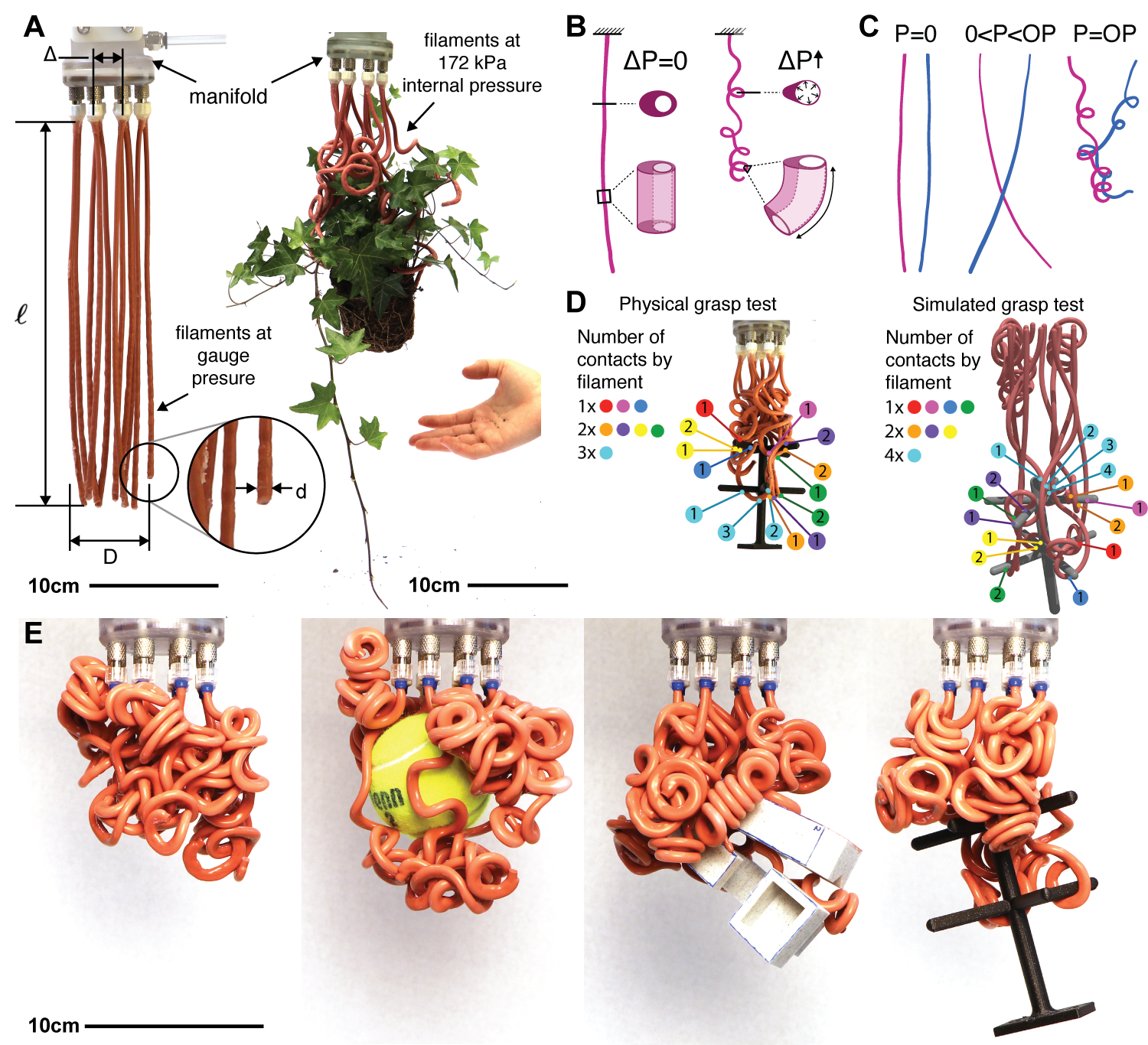}
\caption{ {\bf Fluidically actuated entangling filaments.} ({\bf A})~Photograph of an entangling filament gripper consisting of 12 hollow elastomeric filaments in a resting state and pneumatically actuated around a house plant. ({\bf B})~Schematic of filaments at ambient and increased internal pressure. ({\bf C})~Schematic of the entanglement of two nearby filaments. The filaments are not entangled in their rest state or at low pressures. At low pressures, the filament begins to curl in a plane, and as the internal pressure approaches the operating pressure, (in this case ${172}~{kPa}$,) the filaments bend out of their initial plane and start to entangle with nearby filaments. ({\bf D})~Physical and simulated entanglement examples with contacts between filaments and the object (eight-branch tree) indicated. Contacts are color coded and grouped by individual filaments.
({\bf E})~Photographs of an array of twelve filaments activated by an internal pressure of ${172}~{kPa}$ and entangled around neighboring filaments and various objects.}
\label{fig:F1}
\end{figure}

\clearpage

\begin{figure}[t!]
\centering
\includegraphics[width=.6\textwidth]{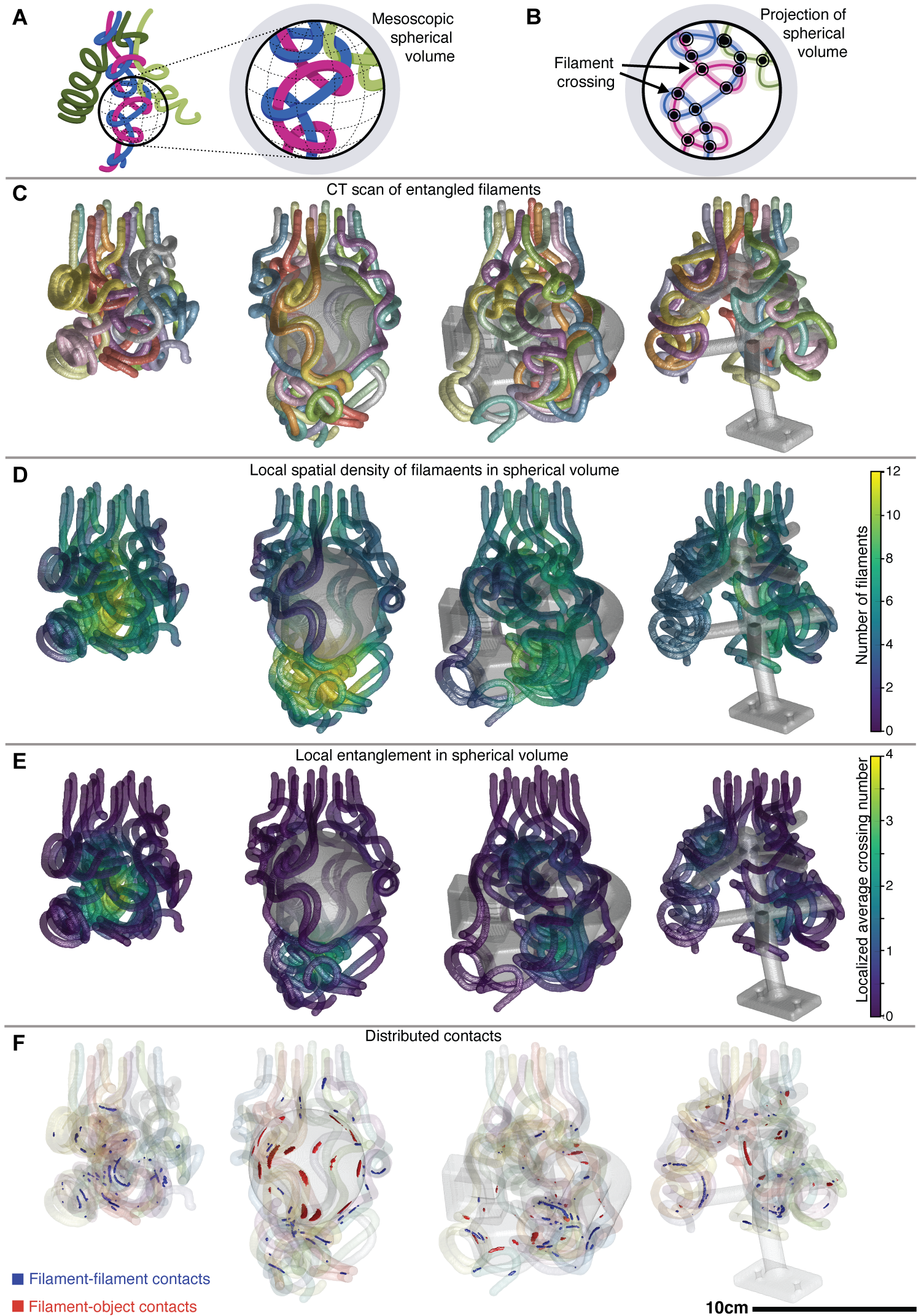}
\caption{ {\bf Spatial distribution of contacts and entanglement.} ({\bf A})~Schematic of four entangled filaments to illustrate a spherical bounding volume used to isolate and evaluate local metrics such as the spatial density of filaments and the localized entanglement density, results of which are presented in Figure~2. ({\bf B})~A projection of the filaments within the spherical bounding volume onto a plane. The number of crossings between filament center lines is used as an indicator of entanglement and averaged over all projection directions to find an average crossing number. ({\bf C})~Micro-computed tomography (micro-CT)-based 3D reconstructions of the entangled filaments and objects used to extract the position and shape of each filament. ({\bf D})~Spatial density of filaments, calculated based on the number of filaments that occur within a spherical bounding volume with a 20 mm diameter. ({\bf E})~Localized average crossing number of the filaments, calculated as an average number of filament crossings over all projections of a spherical bounding volume with a 20 mm diameter. ({\bf F})~3D rendering from \textbf{micro-CT} scans of entangled filaments with filament-filament contacts highlighted in blue and filament-object contacts highlighted in red. The entanglement examples and objects are the same as those shown in Figure~1E and Figure~2C-F.}
\label{fig:F2}
\end{figure}

\clearpage

\begin{figure}
\centering
\includegraphics[width=.6\textwidth]{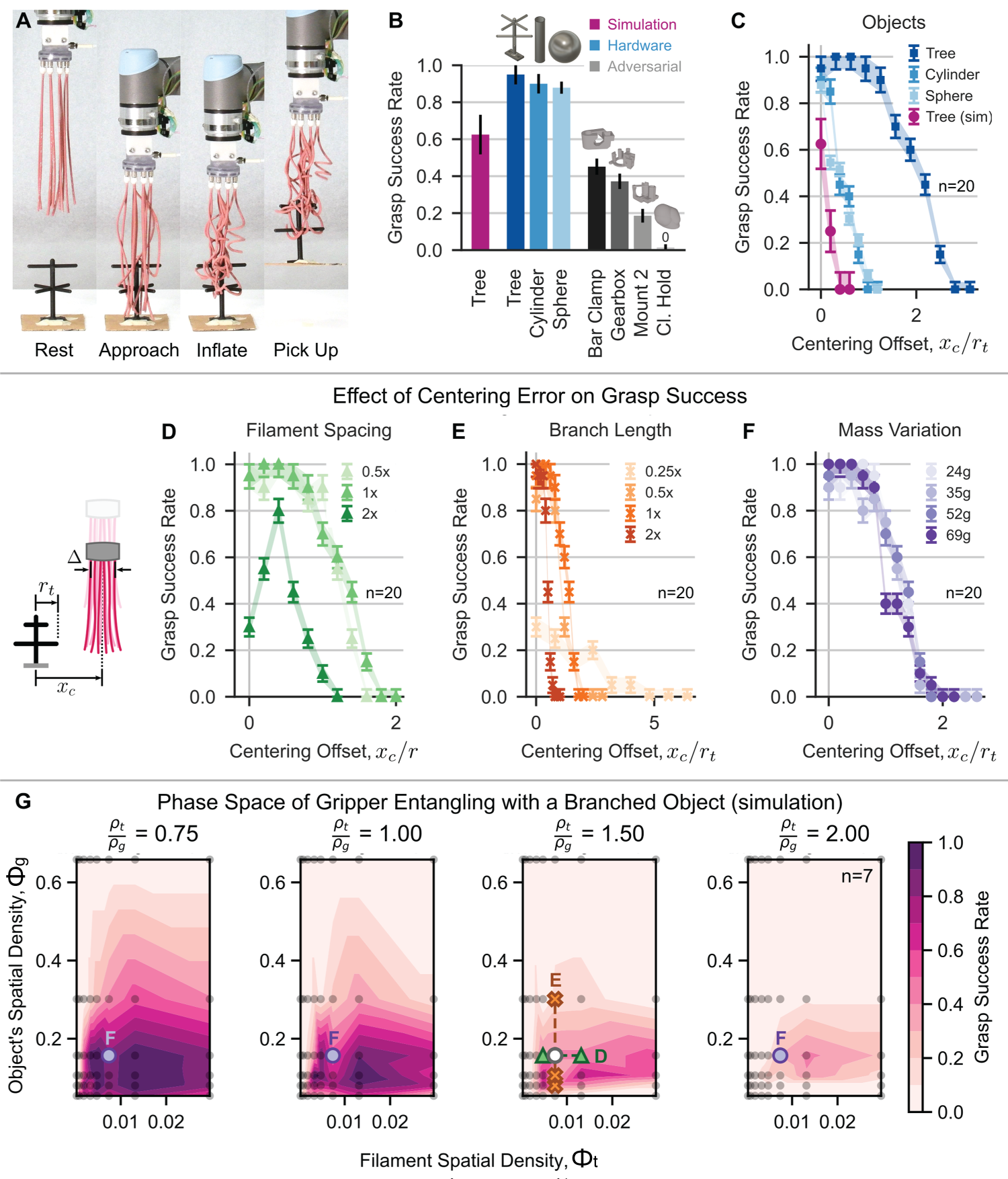}
\caption{{\bf Collective entanglement as a stochastic grasping strategy.} ({\bf A}) Photographs showing a sequence of an entanglement grasping tests conducted with a UR5 robotic arm. ({\bf B}) Success rate of grasping tests performed with various objects that were centered directly below an array of twelve filaments. Objects include the eight-branch tree shown in Figure~3B, a ${38}~{mm}$ diameter cylinder, ${10}~{cm}$ diameter sphere, and four objects from and adversarial object set \cite{mahler2017dexnet2}. Additional object information is included in the supplementary text. All simulated tests were performed with the eight-branch tree. ({\bf C}) Success rate of a grasping test performed with various objects with increasing horizontal offsets between the vertical center line of the array of filaments and the target object. ({\bf D}) Grasping test success rates for a branched object with varying filament spacing and increasing horizontal offsets. ({\bf E}) Grasping test success rates for a branched object with varying spatial density (branch length) and increasing horizontal offsets. ({\bf F}) Grasping test success rates for a branched object with four different weights and increasing horizontal offsets. ({\bf G}) Phase space of grasp success rate predicted by simulations of filaments entangling with the branched test object. Each plot represents a different object weight. Sweeps of object spatial density and filament spatial density are shown within each plot. The phase space locations corresponding to data in D, E, and F are indicated on the plots.}
\label{fig:F3}
\end{figure}

\end{document}